\documentstyle[12pt]{article}
\newcommand{\bea}{\begin{eqnarray}}
\newcommand{\be}{\begin{equation}}
\newcommand{\eea}{\end{eqnarray}}
\newcommand{\ee}{\end{equation}}
\def\nn{\nonumber}

\let\ssection=\section
\renewcommand{\section}{\setcounter{equation}{0}\ssection}

\def\a{\alpha}
\def\b{\beta}

\def\d{\delta}
\def\e{\epsilon}

\def\f{\phi}
\def\g{\raisebox{.4ex}{$\gamma$}}

\def\l{\lambda}

\def\q{\theta}

\def\F{\Phi}

\def\J{\Psi}
\def\L{\Lambda}

\def\P{\Pi}

\def\cc{{\cal C}}
\def\cd{{\cal D}}

\def\cg{{\cal G}}
\def\ch{{\cal H}}

\def\cl{{\cal L}}
\def\cm{{\cal M}}

\def\co{{\cal O}}

\begin{document}
\pagestyle{empty} \large \noindent . \vspace*{5mm} \\
CGPG-94/2-4\\gr-qc/9403002\\February 1994\\revised May 1994\\
\begin{center} \LARGE \bf \vspace*{15mm}
Gravity and Yang-Mills theory: two faces of the same theory?\\
\vspace*{15mm} \large \bf
Subenoy Chakraborty\dag \ddag  and Peter Peld\'{a}n\dag
\footnote{Email address: peldan@phys.psu.edu}\\
\vspace*{5mm} \large
\dag Center for Gravitational Physics and Geometry\\
Department of Physics\\
Pennsylvania State University, University Park, PA 16802, USA\\
\vspace*{5mm}\ddag Department of Mathematics\\
Jadavpur University, Calcutta-700032, India\\
\vspace*{10mm} \Large \bf
Abstract\\
\end{center} \normalsize
We introduce a gauge and diffeomorphism invariant theory on the
Yang-Mills phase
space. The theory is well defined for an arbitrary gauge group with an
invariant bilinear form, it contains only first class constraints, and the
spacetime metric has a simple form in terms of the phase space variables. With
gauge group $SO(3,C)$, the theory equals the Ashtekar formulation of gravity
with a cosmological constant. For Lorentzian signature, the theory is
complex, and we have not found any good reality conditions. In the Euclidean
signature case, everything is real. In a weak field expansion around de
Sitter spacetime, the theory is shown to give the conventional Yang-Mills
theory to the lowest order in the fields. We show that the coupling to a Higgs
scalar is straightforward, while the naive spinor coupling does not work. We
have not
found any way of including spinors that gives a closed constraint algebra.
For gauge group $U(2)$, we find a
static and spherically symmetric solution.\\
PACS: 04.20.Fy, 04.20.+h

\newpage \pagestyle{plain}
\section{Introduction}\label{intro}
The Ashtekar formulation of Einstein gravity is a canonical description on
the $so(3)$ Yang-Mills phase space \cite{Ash1}.
 This formulation has received a lot of
attention mainly due to the fact that the constraints in the theory have a
very simple form, making it possible to restart the old
attempts at a canonical quantization of gravity. There is, however, another
aspect of the Ashtekar's variables that make them very interesting already at
the classical level. Merely the fact that gravity can be so
successfully formulated in
terms of Yang-Mills type variables, could be an indication that there exist a
unified description of gravity and Yang-Mills theory.
Furthermore, there are a few puzzling features of the Ashtekar's variables
that could be the clues that lead us to the correct theory.
 For instance: 1. Why does the theory has to
be complex for Lorentzian signature? 2. Why do we only need half the Lorentz
group in the construction? 3. How come the coupling to Yang-Mills theory seems
more unnatural than the other matter couplings?

It could happen that the answer to the two first questions are the same:
perhaps the other half of the Lorentz group is tied up serving as an
internal symmetry group for some other gauge interaction ($su(2)$), and in
order to
find the pure gravity theory, we need to split $so(1,3)\cong su(2,C)\oplus
su(2,C)$ which necessarily means introducing complex fields. (See
\cite{KacMoody} for a treatment of an $so(1,3)$ Ashtekar formulation.) \\ \\
In this paper, we will present a generalization of the Ashtekar formulation.
It is a generalization in the sense that we now can use an arbitrary gauge
group as internal symmetry group. The model contains only first class
constraints and the spacetime metric has a simple form in terms of the
basic fields. There are, however, several reasons that
make us believe that this is {\it not}
the unified theory we are looking for. For
instance, we still need complex fields in order to get Lorentzian
signature on the metric. Moreover, since the theory seems to work perfectly
alright for an arbitrary gauge group, the theory does not tell us anything
about
what gauge symmetries we should have. With this perspective, the theories
described in \cite{KacMoody} appear to be much more interesting ones, since
they seem to give Lorentzian signature with real fields, and the construction
singles out specific gauge groups. So although the examples chosen in
\cite{KacMoody} were shown to describe nothing more than complex Einstein
gravity, there might still exist other gauge groups that, by using the
construction in \cite{KacMoody}, give physically interesting unified
theories.
There is,
however, another reason that, so far, makes the construction presented
here superior to the one used
in \cite{KacMoody}: the theory presented in this paper gives the standard
Yang-Mills theory when we expand around de Sitter spacetime for weak fields.
Weak fields here means; weak compared to the energy scale set by the
dimension-full constant in the theory: the cosmological constant. It is
actually a necessary requirement, in this type of theories, that one has to
have a dimension-full constant included.
The reason is that in these theories the metric and the Yang-Mills
"electric-field" are roughly speaking comparable objects, and since these two
fields have different dimensions, we need a dimension-full constant for a
rescaling of one of the fields\footnote{Without a dimensionfull constant one
would have to use the Planck constant -- already at the classical level --
in order to be able to unify a metric-type variable and a YM-type electric
field. Also, without a cosmological constant term in the Hamiltonian, it is
hard to see how the conventional Einstein-YM Hamiltonian could be produced
in the weak field limit of the theory.}.
Moreover, when doing a weak field expansion of
the theory (to compare it to conventional Yang-Mills theory on
approximately Minkowskian spacetime) this constant will be the natural
energy scale to define what is meant by "weak fields".
In the theory we study in this paper, this
constant is the cosmological constant, which from cosmological considerations
normally is considered to correspond to a very low energy-density. In
practice this means that we will only get the conventional Yang-Mills theory
for fields that are weak w.r.t the cosmological constant energy-scale. This
seems to be a serious drawback for any unified description where the energy
scale is set by the cosmological constant. Thus, if we believe in
this kind of unification -- with the cosmological constant -- we would expect
to get corrections to {\it e.g} Maxwell's theory already for very low field
strengths (that is if we isolate the experiment from all background energy
densities, since a slowly varying background energy density will manifest
itself as a cosmological constant, and thereby increasing the relevant energy
scale). There is, however, another possibility: by instead introducing a
dimension-full constant that correspond to a very high energy density -- or
very short length -- we would not get into conflict with cosmology, and
Yang-Mills theory would be correct up to much higher energies. This is
actually what one does in string theory when one introduces a classical
string-tension that corresponds to a very short length scale. To do this in
practice, in a unified theory of the type discussed in this paper, one
could modify the Hamiltonian constraint by adding terms that are of higher
order in the magnetic field.\\ \\
The outline of the paper is: in section \ref{Gav} we present the problem of
generalizing the Ashtekar formulation, and in section \ref{arb} we introduce
the explicit model we study in this paper, do a constraint analysis and give
a short discussion of the problem of finding good reality conditions.
In section \ref{weak}, we do a weak field expansion of the Hamiltonian
around de Sitter spacetime, and show that the conventional YM
Hamiltonian is contained in this theory. Section
\ref{matter} shows that it is straightforward to include a Higgs scalar field
in the general theory, while spinor fields seem to be problematic to
incorporate into
the unified formulation.
We derive an explicit static and spherically symmetric solution to
the $U(2)$ theory in section \ref{solution}. In appendix A, we derive the
Hamiltonian for the conventional Yang-Mills theory,
and finally in appendix B, we introduce dimensions and units in the
theories.\\ \\
Throughout this paper, we try to give all conventions needed in the
relevant section. As a general rule, our index conventions are: $\a ,\b , \g,
....$ spacetime indices, $a, b, c, ...$ spatial indices, $i, j, k, ...$
$so(3)$ indices in the vector representation, $I, J, K, ...$ gauge indices
(other than $so(3)$) in the vector representation. $A, B, C,....$ are used
both as Yang-Mills gauge indices in the vector representation as well as
$su(2)$ spinor indices.

Also, compared to for instance \cite{thesis}, we have rescaled the Ashtekar
variables according to: $A_{ai}\rightarrow -iA_{ai}$, $E^{ai}\rightarrow
iE^{ai}$. These conventions are the ones that minimizes the number
of complex "i's" in the theory. (Actually, for pure gravity without a
cosmological constant, {\it no}
explicit "i" appears in the formulation.) These
conventions are also the most natural from the unification point of view.

\section{Generalizing Ashtekar's variables}\label{Gav}
In an attempt to find a unified theory of gravity and Yang-Mills theory (YM),
 on Yang-Mills phase space, a natural starting point is the
generalization of Ashtekar's variables to other gauge groups. The Ashtekar
formulation is a canonical description of (3+1)-dimensional gravity on the
$SO(3,C)$ Yang-Mills phase space \cite{Ash1}, \cite{thesis}. If
the unified theory
 is found from a generalization of the Ashtekar Hamiltonian, we are
guaranteed that the pure gravity theory will appear when we remove the YM
part of the theory. That is, when we choose the gauge group to be $SO(3,C)$
again. But perhaps this is a too naive expectation, perhaps we instead should
look for a more sophisticated construction were the pure gravity part is more
non-trivially embedded. This would for instance be the case if we, in the
formulation of the model, used special objects, features etc that only exist
for "the unified gauge group". See {\it e.g.} \cite{KacMoody} for examples of
that.
 for examples of that.
In this paper, we examine the naive construction.\\ \\
The Ashtekar Hamiltonian for pure gravity with a cosmological constant is:
\bea \ch_{tot}&=& N \ch + N^a \ch_a + \L _i \cg ^i \label{1} \\
\ch&:=&-\frac{1}{4}
\e_{abc}\e_{ijk}E^{ai}E^{bj}(B^{ck}+\frac{2i\l}{3}
E^{ck})\approx 0 \label{2} \\
\ch _a&:=&\frac{1}{2}\e_{abc}E^{bi}B^c_i\approx 0 \label{3} \\
\cg _i&:=&\cd _a E^a _i=\partial _a E^a _i + f_{ijk}A_a^j E^{ak}\approx 0
\label{4}
\eea
The index-conventions are: $a, b, c, .....$ are spatial indices on the three
dimensional hyper surface, and $i, j, k, .......$ are $SO(3,C)$ gauge-indices
in the
vector representation. Gauge-indices are raised and lowered
with an invariant bilinear form of the Lie-algebra (the "group-metric"),
here chosen
to be the Cartan-Killing form, $\d _{ij}$. The basic conjugate fields are
$A_{ai}$ and
$E^{bj}$, which satisfy the fundamental Poisson bracket: $\{
A_{ai}(x),E^{bj}(y)\}=\d ^b _a \d ^j _i \d ^3 (x-y)$. $A_{ai}$ is a gauge
connection, and $E^{ai}$ is often referred to as the "electric field", and it
equals the densitized triad, in a solution. The other
fields in the theory, $N, N^a, \L _i$ are Lagrange multiplier fields whose
variations impose the constraints $\ch, \ch _a$ and $\cg_i$. $B^{ai}$ is the
"magnetic field": $B^{ai}:=\e^{abc}F^i_{bc}=\e^{abc}(2 \partial _b A^i_c +
f^i{}_{jk}A_b^jA_c^k)$, and $\l$ is the cosmological constant. Finally, $\e
_{ijk}$ and $\e _{abc}$ are the totally antisymmetric epsilon-symbols:
 $\e _{123}:=1$ for both of them.\\ \\
The reason why it is a non-trivial task to generalize this Hamiltonian
to other gauge groups can be found in the $\e _{ijk}$ in the Hamiltonian
constraint, $\ch $. If one sees this object as an epsilon-symbol, it is only
well-defined for three dimensional Lie-algebras. If one instead sees this $\e
_{ijk}$ as the structure constant of $SO(3)$ (which it is, with proper
normalization), the introduction of other gauge groups will not give a closed
constraint algebra for the constraints given in (\ref{1})-(\ref{4}). The
problem is the Poisson bracket $\{ \ch ,\ch \}$ which will not weakly vanish
for arbitrary gauge groups. So, the theory would need additional constraints,
possibly of second class, and that would further complicate the attempts at a
canonical quantization of it.

To evade these difficulties, we see three different strategies: {\bf I}.
eliminate the epsilon completely from $\ch $, {\bf II}. generalize the
epsilon-symbol to be well-defined for higher dimensional gauge groups as
well, or {\bf III}. replace the $\e _{ijk}$ with $f_{ijk}$, the structure
constant of the Lie-algebra, and use only gauge groups that gives a closed
constraint algebra.\\ \\
{\bf I}. This method has been described in \cite{arbgg}. One multiplies the
Hamiltonian constraint, $\ch $ with a scalar function containing an $\e
_{ijk}$, and then uses the $\e -\d $ identity to remove all $\e 's$ from the
formulation. The resulting constraints will in general give a closed algebra
for arbitrary gauge groups.\\ \\
{\bf II}. The generalization of the $\e _{ijk}$ to higher dimensional gauge
groups can be done by using the three-dimensionality of the underlying
 hyper surface.
Since there exist an totally antisymmetric tensor density $\e _{abc}$ on the
hyper surface, one can easily construct a Lie-algebra valued $\e _{ijk}$ as
follows:
\be\e_{ijk}({\scriptstyle
V^{dl}}):=\frac{\e_{abc}V^{a}_iV^b_jV^c_k}{\sqrt{det(V^{dl}V^e_l)}}
\label{eps}
\ee
where $V^{ai}$ is a trio of Lie-algebra valued vector fields. Note that
this method actually is contained in {\bf I.}\\ \\
{\bf III}. To use this method, one needs a Lie-algebra whose structure
constants satisfy a certain type of identity. The only Lie-algebras we
know of that can be used here, are $so(4), so(1,3), so(2,2), su(2), so(3),
so(1,2)$ and the Kac-Moody affine extension of these. See \cite{KacMoody}
 for an example of a theory based on the Kac-Moody extension of $so(1,3)$.\\
\\
In this paper, we will study a model based on a generalization of type {\bf
II} (or {\bf I}).

\section{The arbitrary gauge group theory}\label{arb}
In this section, we will study one Hamiltonian that can be found from the
Ashtekar Hamiltonian (\ref{1})-(\ref{4}) by using method {\bf II} (or {\bf
I}) in the previous section. First, we describe the Hamiltonian, valid for
arbitrary gauge groups, then we calculate the constraint algebra and show
that the theory contains only first class constraints, and finally, we give
the geometrical interpretation of the fields. There is also a short
discussion of the (absence of) reality conditions.

\subsection{The Hamiltonian}\label{Hami}
We start by generalizing the Ashtekar Hamiltonian using method {\bf II}, and
choosing the vector fields $V^{ai}$ to be $E^{ai}$. There is nothing unique
about this choice, and the reason for it is that we know that the standard
gravity-Yang-Mills coupling is of sixth order in the electric fields, meaning
that we need $V^{ai}$ to be linear in $E^{ai}$ if we should recover the
standard coupling in this theory. Then, $V^{ai}=E^{ai}$ is the simplest
choice. Thus, the only change from (\ref{1})-(\ref{4}) is that $\ch $ now
becomes:
\be \ch:=-\frac{1}{4}
\e_{abc}\e_{IJK}({\scriptstyle E^{dL}})E^{aI}E^{bJ}(B^{cK}+\frac{2i\l}{3}
E^{cK})\label{uniH}\approx 0
\ee
where
\be \e_{IJK}({\scriptstyle
E^{dL}}):=-\frac{\e_{abc}E^{a}_IE^b_JE^c_K}{\sqrt{det(E^{dL}E^e_L)}}.
\label{epsE}
\ee
There is also the invisible change that the gauge indices $I,J,K,...$ now
take values $1,2,...N$, where $N$ is the dimension of the Lie-algebra. Also,
we suppose that there exist an invariant bilinear form of the Lie-algebra
that can be used to contract the gauge indices. (For the semi-simple
classical groups, this form can always be chosen to be the Cartan-Killing
form.) Note that the definition of $\e_{IJK}({\scriptstyle
E^{dL}})$ requires a non-zero $det(E^{aI}E^b_I)$, which later will be shown
to correspond to a non-degenerate metric. Since we will have to allow the
fields to be complex, we really get an sign-ambiguity from the square-root
in the $\e_{IJK}({\scriptstyle
E^{dL}})$. This sign-ambiguity just corresponds to a sign change of the
total spacetime metric. If one wants to avoid this inconvenient
square-root, one could instead use $\e _{abc}f_{IJK}E^{aI}E^{bJ}E^{cK}$ as
denominator. This would give an inequivalent $\e_{IJK}({\scriptstyle
E^{dL}})$, which, however, coincides with the conventional $\e _{IJK}$
for $SO(3)$. We will choose (\ref{epsE}) as our definition. \\ \\
Before we go on and study the Hamiltonian, a few words should be said about $
\e_{IJK}({\scriptstyle E^{dL}})$ itself. This object has the following
important properties: it is totally antisymmetric, it reduces to the
conventional, constant epsilon-symbol for three dimensional algebras, and it
satisfies an "$\e - \d $" identity on the three dimensional subspace (in the
Lie-algebra) spanned by $E^{aI}$. The antisymmetricity follows from
antisymmetric property of $\e _{abc}$, and by using the three dimensional
identities
\bea \e _{abc}E^{aI}E^{bJ}E^{cK}&=&det(E^{dL})\e^{IJK} \\
det(E^{aI}E^b_I)&=&(det(E^{aI}))^2 \eea
it follows that $\e_{IJK}({\scriptstyle
E^{dL}})=\pm \e_{IJK}$ for three-dimensional algebras. To give the "$\e -\d
$" identity, we first need to define a few objects:
\bea q^{ab}&:=&-E^{aI}E^b_I \\
q_{ab}&:=&\frac{1}{2 \mid q^{ef}\mid}\e _{acd}\e _{bef}q^{ce}q^{df}\Rightarrow
q_{ab}q^{bc}=\d _a^c \\
E_{a}^I&:=&-q_{ab}E^{bI} \Rightarrow E_{aI}E^{bI}=\d ^b _a\\
\tilde{\d}^I_J&:=&E_{aJ}E^{aI} \eea
The $\tilde{\d}^I_J$ is a projection operator that projects down to the three
dimensional space spanned by the $E^{aI}$. With these definitions, the "$\e
-\d $" identity becomes
\be \e_{IJK}({\scriptstyle E^{dL}})\e^{LMN}({\scriptstyle
E^{dL}})=\tilde{\d}^{[L}_I\tilde{\d}^M_J\tilde{\d}^{N]}_K \ee
where the square brackets denotes antisymmetrization of all indices inside,
 without factors.

\subsection{Constraint analysis}\label{constraint}
Returning to the Hamiltonian, one needs to check that the theory is
consistent in the sense that an initial field configuration that satisfy all
constraint will continue to do so. That is the same as requiring the time
evolution of the
constraints to be weakly vanishing (zero, modulo all constraints). And for
Hamiltonians which only consist of constraints, this corresponds to checking
the constraint algebra. If all constraints are first class, i.e all Poisson
brackets between constraints result in linear combinations of constraints,
the time-evolution of the constraints will weakly vanish.

In calculating the constraint algebra in a gauge and diffeomorphism invariant
theory, it is often convenient to first identify the generators of gauge
transformations and diffeomorphisms on the hyper surface. Since these
generators have a simple action on all gauge and diffeomorphism covariant
objects, it is a simple task to calculate all Poisson brackets containing
these generators. For theories formulated on Yang-Mills phase space it is
well known that the generator of gauge transformations is the Gauss law
constraint, $\cg _I$, and that the generator of spatial diffeomorphisms is
given by $\tilde{\ch }_a:=\ch _a -A_a^I\cg _I$, where $\cg _I$ and $\ch _a$
are given in (\ref{3}) and (\ref{4}). (This is true in arbitrary spacetime
dimensions $>2$, and for arbitrary gauge groups. See e.g \cite{thesis}.)
To prove this, consider the transformations generated on the basic fields
$A_a^I$ and $E^{a}_I$. (Square brackets here denote smearing over the hyper
surface: $\cg _I[\L ^I]:=\int _{\Sigma}d^3x \L ^I\cg _I$.)
\bea \d ^{\cg ^I}E ^{aI}&:=&\{E ^{aI},\cg ^J[\L_J]\}=f^I{}_{JK}\L^JE ^{aK}
\label{6.3} \\
\d ^{\cg ^I}A _a ^I&=&\{A _a ^I,\cg ^J[\L_J]\}=-\cd _a\L^I \label{6.4} \eea
which shows that $\cg _I$ is indeed the generator of gauge transformations, and
\bea \d ^{\tilde{\ch}_a}E ^{aI}&:=&\{E ^{aI},\tilde{\ch}_b[N^b]\}\!=\!N^b
\partial _b
 E
^{aI} \!-\! E ^{bI}\partial _b N^a \!+\! E ^{aI} \partial _b N^b\!=\!
\pounds _{N^b}E ^{aI}
\label{6.6} \\
\d ^{\tilde{\ch}_a}A _a ^I&=&\{A _a ^I,\tilde{\ch}_b[N^b]\}\!=\!N^b\partial _b
A _a ^I \!+\!
A _{b}^I\partial _a N^b\!=\!\pounds _{N^b}A _a ^I \label{6.7} \eea
prove that $\tilde{\ch }_a$ generates spatial diffeomorphisms. Then it is
clear how any gauge and diffeomorphism covariant function of $A_a^I$ and
$E^{a}_I$ transforms:
\begin{eqnarray}
\delta^{{\cal G}^I [\Lambda_I]} \Phi_{aI}
({\scriptstyle E^a_I,A_b^J})&=&\Lambda^J
\Phi_a^K({\scriptstyle E^a_I,A_b^J}) f_{JKI}\label{3.14} \\
\delta^{\tilde{{\cal H}}_a[N^a]} \Phi_{aI} ({\scriptstyle E^a_I,A_b^J})&=
&\pounds_{N^a}
\Phi_{aI}({\scriptstyle E^a_I,A_b^J}) \end{eqnarray}
Thus, knowing that $\cg _I$ and $\ch$ are gauge-covariant, and that $\cg _I$,
$\tilde{\ch}_a$ and $\ch$ all are tensor densities, the Poisson brackets
containing $\cg _I$ and $\tilde{\ch} _a$ are easily calculated:
\begin{eqnarray}
\{{\cal G}^I [\Lambda_I],{\cal G}^J[\Gamma_J]\}&=&{\cal G}^K[f_{KIJ}\Lambda^I
\Gamma^J]\label{3.15} \\
\{{\cal G}^I [\Lambda_I],\tilde{{\cal H}}_a[N^a]\}&=&{\cal
G}^I[-\pounds_{N^a}\Lambda^I]\\
\{{\cal G}^I [\Lambda_I],{\cal H}[N]\}&=&0\\
\{\tilde{{\cal H}}_a[N^a],\tilde{{\cal H}}_b[M^b]\}&=&\tilde{{\cal H}}_a
[-\pounds_{M^b}
N^a]\\
\{{\cal H}[N],\tilde{{\cal H}}_a[N^a]\}&=&{\cal H}[-\pounds_{N^a}N]
\label{3.15b}
\end{eqnarray}
This leaves only one Poisson bracket left to calculate: $\{ \ch [N],\ch
[M]\}$. This is a rather complicated calculation but it simplifies to note
that the result must be antisymmetric in $N$ and $M$, meaning that it is only
terms containing derivatives of these test functions that survive.
Furthermore, by first checking that $\{det(E^{aI}E^b_I)[N],\ch
[M]\}- (N \leftrightarrow M) =0$, one sees that we only have two contributions
to the end result:
\bea &&\{ \ch [N],\ch
[M]\}=\{B^b_I[{\scriptstyle \frac{N}{4\sqrt{-det(q^{ab})}} \e _{bcd}
\e
_{aef}q^{ce}q^{df}E^{aI}}],E^g_J[{\scriptstyle -\frac{M}{2}\sqrt{-det
(q^{ab})}q_{gh}B^{hJ}}]
\} + \nn \\
 &&\{B^b_I[{\scriptstyle \frac{N}{4\sqrt{-det(q^{ab})}} \e _{bcd}
\e _{aef}q^{ce}q^{df}E^{aI}}],E^{a_1}_J[{\scriptstyle -\frac{M}{2
\sqrt{-det(q^{ab})}}
E^{(e_1K}B^{f_1)}_K\e _{e_1 a_1 b_1}\e _{f_1 c_1 d_1} q^{b_1d_1}E^{c_1J}}]\} -
\nn \\
&& (N\leftrightarrow M) = -\frac{1}{2}(N\partial _cM - M\partial _cN) q^{ce}
\e _{aeb}E^{aI}B^b_I - \nn \\
&& \frac{1}{2det(q^{ab})}(N\partial _cM - M\partial _cN) q^{a_1f}q^{ac_1} \e
_{aef} q^{ce} E^{(e_1K}B^{f_1)}_K \e _{e_1a_1b_1} \e _{f_1c_1 d_1} q^{b_1d_1}
= \nn \\  && \ch _a[-q^{ab}(M\partial _bN-N\partial _bM)] \label{pb} \eea
where we in the last step used $\e _{abc}q^{bd}q^{ce}=det(q^{ab})q_{af}\e
^{fde}$.

Thus the constraint algebra closes and the theory is complete and
consistent, in this sense. Furthermore, the constraints satisfy the general
constraint algebra that is required for any diffeomorphism invariant theory,
with a metric \cite{HKT}. In ref. \cite{HKT} Hojman {\it et al} showed that
in any canonical formulation of a diffeomorphism invariant theory, with a
metric, where one can find a set of constraints $\cc _a$ and $\cc$
generating spatial diffeomorphisms and time-like diffeomorphisms, these
constraints will always satisfy the algebra above, with $\cc
_a=\tilde{\ch}_a$ and $\cc =\ch$. It was also shown, as a consistency
requirement of path-independence of deformations, that the spatial metric has
to appear as a structure function in the Poisson bracket corresponding to
(\ref{pb}) above. In our case, this means that the densitized spatial metric
on the hyper surface is given by
\be q^{ab}:=-E^{aI}E^b_I. \label{metric} \ee
However, note that we, at this point, cannot say anything about the physical
relevance of this metric. Is this the metric that defines the causal
structure of the theory? So far, we do not have any answer to that question,
it should, however, be clear that this is the only consistent choice for a
metric. If it later is shown that this theory allows non-causal propagation
with respect to the light cones of the metric, the theory is
unphysical and must be abandoned.

Now, the constraint algebra gave us the spatial metric. What about the
complete spacetime metric? Knowing the spacetime metric is the same as
knowing how the spatial metric evolves from one hyper surface to another,
which corresponds to the time-evolution of the spatial metric. The following
is true \cite{HKT}
\be \{ q^{ab},\ch [N]\}=-2 N g^{\frac{3}{2}}(K^{ab}-g^{ab}K^c_c) \label{qdot}
\ee where $g^{ab}=\frac{1}{\sqrt{\mid q^{cd}\mid}}q^{ab}$ and $K^{ab}$ is the
extrinsic curvature of the hyper surface embedded in spacetime. This,
together with the conventional expression for the extrinsic curvature in
terms of the spacetime metric, makes it straightforward to find the form of
the spacetime metric in terms of the fields in the theory:
\be \tilde{g}^{\a \b}=\sqrt{-g}g^{\a \b}=\left(\begin{array}{cc}
-\frac{1}{N}& \frac{N^a}{N}\\
\frac{N^a}{N}& -NE^{aI}E^b _I -\frac{N^aN^b}{N}
\end{array} \right) \label{Metric} \ee
With such a simple form of the metric, it is an easy task to put the
Lorentz-signature condition on the basic fields. (In \cite{arbgg} another
generalization of the Ashtekar Hamiltonian was given, but for that model, the
metric had a very complicated form, making it
almost impossible to put the signature condition directly on the basic
fields.) Here, Lorentz-signature corresponds to $q^{ab}$ positive definite.

It is at this stage the need for complex fields appears (for Lorentzian
signature). If $q^{ab}$ is positive definite, $\e _{IJK}({\scriptstyle
E^d_L})$ becomes complex, unless $E^{aI}$ is imaginary. In both cases, the
formulation becomes complex. If one tries to rescale the Hamiltonian by
multiplying with $\sqrt{\mid E^{aI}E^b_I\mid}$, the result is that the
right-hand side of (\ref{pb}) changes so that the spatial metric becomes
$-\mid E^{cK}E^d_K\mid E^{aI}E^b_I$, a metric that by construction cannot
be positive definite for real fields.
For Euclidean signature, $q^{ab}$ is required to
be negative definite, which can be accomplished without ever having to
introduce complex fields. Thus, the Lorentzian case still needs complex
fields regardless of the signature of the "group-metric", and the Euclidean
case works perfectly all right with real fields. This does not, however, mean
that the hope of finding a real (non-complex) unified theory is completely
dead. It may be that there exist another type of modification of the Ashtekar
variables that gives a real theory, perhaps a modification that only works for
 a very special gauge group with
some particular feature. After all, that is precisely what we want; a problem
in the theory that can only be solved by choosing the {\it correct} gauge
group.

\subsection{Reality conditions}\label{reality}

If we want to have a consistent and physically
sensible theory in the Lorentzian case, we must allow the fields to be
complex, and, furthermore, we need good reality conditions that will give us
real physical quantities. The question is then, what objects are physical,
and should therefore always be real in a solution? Normally, one says that the
metric is physical while for instance a Yang-Mills connection is not
considered physical. The reason for this is that the connection is not
gauge-invariant and therefore cannot be a physical measurable object, while
the metric normally is considered to be measurable, in some sense. Also, if
one consider the underlying Lorentzian spacetime to be real (which one
normally does), the invariant proper time $d\tau ^2=g_{\a \b}dx^{\a}dx^{\b}$
will be complex if the metric is not real. However, seen from the point of
view of canonical formulations of theories with local symmetries, the
diffeomorphism symmetry is on the same footing as the gauge-symmetry. And,
according to Dirac, no objects that transform under the transformations
generated by the first class constraint should be considered physical. This
means that only gauge and diffeomorphism invariant objects are to be
considered measurable. And, if one really scrutinizes what is done when the
metric
is "measured", one realizes that the metric always is coupled
to some matter fields or
test-particles, meaning that it is really the diffeomorphism invariant
combination of metric and matter that is measured.

This discussion is included to show that it is not so obvious how to impose
the reality conditions. Of course, the safe way is to require all "electric"
and "magnetic" fields to be real, although that may be a too severe
restriction on the theory. (Actually, in the Ashtekar formulation for pure
gravity, one cannot require the "magnetic" field to be real since it equals
the self dual part of the Riemann tensor, in a solution.)

A more sensible compromise is to say that the metric and all gauge-invariant
combinations of the  Yang-Mills part should be real. But already here
we run into problems; how can one impose these conditions on the
Yang-Mills part without breaking the full symmetry. That is, in the full
theory, where the fields take values in the unified Lie-algebra, there is no
splitting between gravity and the Yang-Mills fields, so how can one
impose
reality conditions that only restricts one of the sectors? And, a unified
theory where the reality conditions can only imposed after the symmetry has
been broken, seems rather artificial.

To us, all these complications and the uglification of the theory due to the
need
for complex fields and reality conditions, indicate that this theory
is not the
unified description of gravity and Yang-Mills theory we are looking for.
Instead, we believe that there probably exists another (real?) unified theory,
possibly constructed with the use of a, more or less, unique gauge group, and
that this formulation could be very close to the model presented here.
Finally, since we do not have any good reality conditions, we cannot really
say that we have a theory for Lorentzian signature; we have a theory for
Euclidean signature or complex gravity.

\section{Weak-field expansion}\label{weak}
A minimum requirement on a unified theory of gravity and Yang-Mills theory is
that the Einstein theory of gravitation and the standard Yang-Mills theory
appear in some limit of the theory. We already know that the unified model,
presented in the previous section, reduces to the Ashtekar formulation for
Einstein gravity with a cosmological constant if the gauge group is chosen to
be $SO(3,C)$. Thus, what is left to check is if the conventional Yang-Mills
theory appears in some limit of the theory, and the natural limit to consider
is weak fields around approximate flat spacetime. That is, we know that
 the $U(1)$ Yang-Mills theory on Minkowski spacetime describes
electromagnetism to a very high accuracy. So, in order not to
get in conflict with that observation, we need to find the $U(1)$ Yang-Mills
theory when we expand the unified theory around approximate Minkowski
spacetime (approximate, since the non-zero cosmological constant forces us to
expand around de Sitter spacetime instead).

To do this expansion, we assume that the unified symmetry is broken down to a
direct-product symmetry: $G^{tot}=SO(3,C)\times G^{YM}$ where $SO(3,C)$ is
supposed to be the gravitational symmetry group, and $G^{YM}$ is the internal
symmetry group for the Yang-Mills part. The gauge indices are; $i, j, k, ...$
denote $SO(3,C)$ indices, while $A, B, C,
....$ are the $G^{YM}$ indices. For instance: $E^{aI}E^{b}_I=E^{ai}E^b_i +
E^{aA}E^b_B$. We do not need to use the explicit de Sitter solution to find
the expansion, it suffices to know that
\be  B^{ai}=-\frac{2i \l}{3} E^{ai} \label{de Sitter} \ee
for the de Sitter solution, in terms of Ashtekar variables. See section
\ref{conv}.
The exact solution will be denoted by a bar on top of the symbols, and the
perturbations around it with lower case letters.
\bea E^{ai}&=&\bar{E}^{ai}+e^{ai} \label{pert} \\
     E^{aA}&=&\bar{E}^{aA}+e^{aA}=e^{aA} \nn \\
     B^{ai}&=&\bar{B}^{ai}+b^{ai}=-\frac{2i \l}{3}\bar{E}^{ai}+b^{ai} \nn \\
     B^{aA}&=&\bar{B}^{aA}+b^{aA}=b^{aA} \nn \\
     N&=&\bar{N}+n \nn \\
     N^a&=&\bar{N}^a + n^a = n^a \nn \\
     \L _I&=&\bar{\L}_I + \l _I \nn \eea
Here, we have used (\ref{de Sitter}) and the fact that the Yang-Mills fields
vanishes
in the de Sitter solution. We have also picked a coordinate-system where the
shift vector $\bar{N}^a$ vanishes.
Now, weak-field expansion here means: $e^a_i \ll \bar{E}^a_i$,
$b^a _i \ll \l \bar{E}^a_i$, $e^{aA}e^b_A \ll
\bar{q}^{ab}:=-\bar{E}^{ai}\bar{E}^b_i$ and $e^{aA}b^b_A \ll \l \bar{E}^{ai}
\bar{E}^b_i$. Since we are mainly interested in the Yang-Mills part, we do
not write out the pure gravity perturbation (although, it is a straightforward
task to use the above expansion to get the gravitational part as well):
\bea N\ch&=&\bar{N}\ch ^{(2)}_{YM}+N\ch _{GR}+ \co (e^4,b^4,...)
\nn \\
N^a\ch _a&=&N^a\ch _a^{GR} + \co (e^4,b^4,....) \label{10} \\
\L ^I\cg _I&=&\l^A\cg ^{(2)YM}_A+\L ^i\cg _i^{GR} + \co
(e^4, b^4,....) \nn \eea
where
\bea \bar{N}\ch ^{(2)}_{YM}&=&\frac{\bar{N}}{4}\frac{1}{\sqrt{-\mid
\bar{q}^{ab}\mid}}(\e _{abc}\e _{def}\bar{q}^{ad}\bar{q}^{be}e^{cA}(b^f_A +
\frac{2 i \l}{3} e^f_A)) \label{20} \\
\l^A\cg ^{(2)YM}_A&=&\l ^A \cd _a e^a_A=\l ^A(\partial _a e^a_A +
f_{ABC}a_a^Be^{aC}) \nn \eea
and the $\co (e^4, b^4,....)$ term includes all the higher order terms. $\mid
\bar{q}^{ab}\mid$ is the determinant of the spatial metric
$\bar{q}^{ab}:=-\bar{E}^{ai}\bar{E}^b_i$.
These expressions are to be compared to the conventional Yang-Mills
total Hamiltonian on any fixed background \cite{ART}, \cite{Rom}:
\be \ch ^{conv}_{tot}=\frac{N}{4 \sqrt{\mid q^{ab}\mid}} \e _{abc}\e
_{def}q^{ad}q^{be}(e^{cA}e^{f}_A+\frac{1}{4}b^{cA}b^f_A) +N^a\frac{1}{2}\e
_{abc}e^{bA}b^c_A + \l ^A\cd _a e^a _A
\label{30} \ee
To get exact agreement, we perform a canonical transformation in the
unified theory: $\tilde{e}^{aA}:=e^{aA}-\frac{3 i}{4 \l} b^{aA}$, and $a_{aA}$
unchanged. With this,
(\ref{20}) becomes:
\be \bar{N}\ch ^{(2)}_{YM}=\frac{-\bar{N}}{2 \sqrt{\mid \bar{q}^{ab}\mid}}
\e _{abc}\e_{def}\bar{q}^{ad}\bar{q}^{be}(\frac{\l}{3}\tilde{e}^{cA}\tilde{e}
^f_A + \frac{3}{16\l}b^{cA}b^f_A) \label{40} \ee
where we see that the physical Yang-Mills fields are:
\bea e^{aA}_{phys}&=&\sqrt{\frac{2 \l}{3}}\tilde{e}^{aA} \label{50} \\
 b^{aA}_{phys}&=&\sqrt{\frac{3}{2\l}}b^{aA} \nn \eea
and that the unified theory (\ref{1}) reproduces the conventional
Yang-Mills theory to lowest order.
This means that in a weak field expansion around de Sitter
spacetime, the rescaled Yang-Mills fields (\ref{50}) will be governed by the
Yang-Mills equations of motion. For a $U(1)$ Yang-Mills field, we know that
these equations of motion are Maxwell's equations which are very well
experimentally confirmed. So the question is, for what energy scales does
this unified theory predict significant corrections to the Maxwell's
equations? For a very small $\l$ (since $\l$ has dimension inverse length
square, we really mean; on length scales where $\l r^2 \ll 1$) we know that
the de Sitter metric is approximately the Minkowski metric. So, $ \bar{E}^{ai}
\bar{E}^b_i \approx \d ^{ab}$ in cartesian coordinates. The weak field
expansion is then good for

\be e^{aA}_{phys}e^b_{A,phys}\ll \frac{2 \l}{3}\d ^{ab} \label{70} \ee
With an experimental upper bound on $\l$ of $10^{-62}$ $m^{-2}$ \cite{300}
this restricts the electric field to be much weaker than $10^{-4}V/m$! This
seems to be a severe problem for this theory: it predicts large corrections
to Maxwell's equations already for rather modest field strengths. Note
however that the cosmologically constant here really just is a representative
for any slowly varying background energy density. This means that in an
experiment in a lab here on earth we must include in $\l$ all the
contributions coming from e.g thermal energy. (In room
temperate air, the heat energy-density is about $10^{-40}$ $m^{-2}$ in natural
units, which means that the restriction on the electric field increases to
$10^8 V/m$.)

\section{Matter couplings}\label{matter}
Although it is an interesting achievement by itself to find a consistent
unification of gravity and Yang-Mills type interactions, the theory is not of
much use if it doesn't allow the introduction of matter couplings (at least
spinors). In this section we show that the introduction of a Higgs type scalar
field (for arbitrary gauge group) is straightforward, while spinors cannot be
included
by a simple generalization of the conventional spinor coupling.
The scalar field coupling is found by a trivial generalization
of the standard scalar field coupling in the Ashtekar formulation \cite{ART}.
We
show that the constraint algebra continues to be closed without the
introduction of further constraints. Moreover, the simple form of the
spacetime metric (\ref{Metric}) still holds. We have not looked for/found
any reality conditions here either.

\subsection{Scalar field}\label{scalar}
As mentioned above, the coupling to the scalar field is found by trivially
generalizing the scalar field coupling in the conventional Ashtekar
formulation
\cite{ART}. Here, we consider the scalar field to take values in the vector
representation of the unified Lie-algebra. If one wants to treat scalar
fields which transforms as scalars under the internal symmetry group as well,
this is easily accomplished by simply dropping the internal index on the
scalar field. Here is the Hamiltonian.
\bea H ^{tot}&=&\int _{\Sigma}d^3x\left (N(\ch ^G+\ch ^{\Phi})+N^a(\ch _a^G+
\ch _a^{\F})+\L ^I(\cg
_I^G + \cg _I^{\F})\right ) \label{scalartot} \\
\ch ^G&:=&-\frac{1}{4}
\e_{abc}\e_{IJK}({\scriptstyle E^{dL}})E^{aI}E^{bJ}(B^{cK}+\frac{2i\l}{3}
E^{cK}) \\
\ch ^{\F}&:=&\frac{1}{2} \P ^I\P _I +\frac{1}{2} q^{ab}(\cd _a \F ^I)(\cd
_b\F _I)+\frac{1}{2}m^2 \F ^I\F _I \sqrt{\mid q^{ab}\mid} \label{scalarH} \\
\ch ^G_a&:=&\frac{1}{2}\e_{abc}E^{bI}B^c_I \\
\ch _a^{\F}&:=& \P ^I\cd _a \F _I=\P ^I(\partial _a \F _I + f_{IJK}A_a^J\F
^K) \label{scalarV} \\
\cg ^G_I&:=&\cd _a E^a _I=\partial _a E^a _I + f_{IJK}A_a^J E^{aK} \\
\cg _I^\F &:=&f_{IJK}\F ^J\P ^K \label{scalarG} \eea
The $\e _{IJK}({\scriptstyle E^{dL}})$ is given in (\ref{epsE}), and
$q^{ab}:=-E^{aI}E^b_I$. The fundamental Poisson bracket for the scalar field
$\F ^I$ and its momenta $\P _I$ is: $\{ \F ^I(x),\P _J(y)\}=\d ^I_J\d
^3(x-y)$.

Now, to calculate the constraint algebra, we again use the shortcut described
in section \ref{constraint}; first identify the generators of gauge
transformations and spatial diffeomorphisms, then calculate all Poisson
brackets containing these generators using the transformation-properties of
the constraints.

First, consider the transformations generated by the total Gauss law
constraint, $\cg _I^{tot}:= \cg _I^G + \cg _I^{\F}$. (We also define
$\ch _a^{tot}$ and $\ch ^{tot}$ in an analogous way) Since the matter part
$\cg _I^\F$ does not depend on $A_{aI}$ or $E^{aI}$, the transformations of
them are unaltered, and given in (\ref{6.3}) and (\ref{6.4}). For $\F ^I$ and
$\P _I$, we get
\bea \d ^{\cg _I^{tot}}\F ^I&:=&\{\F ^I,\cg _J^{tot}[\L ^J]\} = f^{IJK}\L _J
\F _K \label{deltafi} \\
 \d ^{\cg _I^{tot}}\P _I&:=&\{\P _I,\cg _J^{tot}[\L ^J]\} = f_{IJK}\L ^J\P ^K
\label{deltapi} \eea
which shows that $\cg_I^{tot}$ generates gauge transformations on all the
basic fields. Then, we define $\tilde{\ch}_a:=\ch _a^{tot} -A_{aI}\cg
^I_{tot}$. The matter part of this constraint is manifestly independent of
$E^{aI}$, and the terms containing both $A_{aI}$ as well as matter fields
cancel, implying that the transformations on $A_{aI}$ and $E^{aI}$ are still
given by (\ref{6.6}) and (\ref{6.7}). The action on the matter fields are
\bea \d ^{\tilde{\ch}_a^{tot}}\F^I&:=&\{\F ^I,\tilde{\ch}_a^{tot}[N^a]\} =
N^a\partial _a \F^I=\pounds _{N^a}\F ^I \label{delta2fi} \\
\d ^{\tilde{\ch}_a^{tot}}\P _I&:=&\{\P _I,\tilde{\ch}_a^{tot}[N^a]\} =
\partial
_a(N^a\P _I)=\pounds _{N^a}\P _I \label{delta2pi} \eea
which again shows that $\tilde{\ch}^{tot}_a$ is the true generator of spatial
diffeomorphisms. With these results, we do not even have to write out the
Poisson brackets containing $\cg _I^{tot}$ and $\tilde{\ch}_a^{tot}$. The
result is already given in (\ref{3.15})-(\ref{3.15b}). The last Poisson
bracket is $\{\ch ^{tot}[N],\ch ^{tot}[M]\}$, and to do this calculation one
should again notice that only terms containing derivatives of $N$ and $M$
will survive. Thus, an inspection of the Hamiltonian constraint gives that
the only terms that could give a non-zero result are: $\{ \ch ^G[N],\ch
^G[M]\}$, $\{\ch ^G[N],\frac{1}{2}q^{ab}\cd _a\F ^I\cd _b \F _I[M]\} - (N
\leftrightarrow M)$ and $\{\frac{1}{2}\P ^I\P _I[N],\frac{1}{2}q^{ab}\cd _a
\F ^I\cd _b \F _I[M]\} - (N
\leftrightarrow M)$. The final result is
\be \{\ch ^{tot}[N],\ch ^{tot}[M]\}=\ch _a^{tot}[q^{ab}(N\partial _bM-M
\partial
_bN)], \label{HH} \ee
showing that the constraint algebra is closed. Thus, the theory described by
(\ref{scalartot}) is complete and consistent. From (\ref{HH}) it is also
clear that the spatial metric is still given by $q^{ab}=-E^{aI}E^b_I$.
Furthermore, since the total structure of the algebra is completely similar
to the matter-free case, it follows that the entire spacetime metric is again
given by (\ref{Metric}).

\subsection{Spinor field}\label{spinor}

We have not managed to find a spinor coupling that gives a closed constraint
algebra
for the unified theory. Here, we just want to show what the problem is of using
the
obvious generalization of the conventional spinor coupling.

Since we know the form of the
conventional gravity-spinor coupling in Ashtekar's variables \cite{ART}, the
obvious
first attempt to include spinors into the unified theory, is to generalize that
coupling. Thus, we replace the gravitational part of the Hamiltonian constraint
with
(\ref{uniH}), and replace the Pauli-matrices -- the spin-$\frac{1}{2}$
representation of $so(3)$ -- with the representation $T ^I_A{}^B$ for the
unified
gauge Lie-algebra.  \\ \\
The Hamiltonian is
\bea H ^{tot}&=&\int _{\Sigma}d^3x\left (N(\ch ^G+\ch ^{Sp})+N^a(\ch _a^G+
\ch _a^{Sp})+\L ^I(\cg
_I^G + \cg _I^{Sp})\right ) \label{spinortot} \\
\ch ^G&:=&-\frac{1}{4}
\e_{abc}\e_{IJK}({\scriptstyle E^{dL}})E^{aI}E^{bJ}(B^{cK}+\frac{2i\l}{3}
E^{cK}) \\
\ch ^{Sp}&:=&-\sqrt{2}E^a_IT ^{I}_A{}^B\P ^A\cd _a \l _B
 \label{spinorH} \\
\ch _a&:=&\frac{1}{2}\e_{abc}E^{bI}B^c_I \\
\ch _a^{Sp}&:=& \P ^A \cd _a \l _A  \label{spinorV} \\
\cg _I&:=&\cd _a E^a _I=\partial _a E^a _I + f_{IJK}A_a^J E^{aK} \\
\cg _I^{Sp} &:=&-\frac{1}{\sqrt{2}}T _{IA}{}^{B}\P ^A\l _B
 \label{spinorG} \eea
The fundamental spinor field, $\l _A$ and its conjugate momenta,
$\P ^B$ satisfy the basic Poisson brackets:
\be \begin{array}{ll} \{\P ^B(x),\l _A(y)\}=\d ^B_A\d ^3(x-y) &
\{\l _A(x),\P ^B(y)\}=\d ^B_A\d ^3(x-y) \end{array} \ee
Here, $A, B, C, ... $ are spinor indices in the representation space
of the unified Lie-algebra, $I, J, K, ... $ denote Lie-algebra
indices in the vector
representation, and $T_{IA}{}^B$ is a representation of the unified
Lie-algebra.
The spinors are taken to be grassman odd: {\it e.g }$\l_A \J _B=-\J _B \l_A$.
The action of the covariant
derivative on the spinors is:
\bea \cd _a\l _A&=&\partial _a\l _A -\frac{1}{\sqrt{2}}A_{aI}T ^I_A{}^B\l _B
\label{covder}\\
\cd _a\P ^A&=&\partial _a\P ^A +\frac{1}{\sqrt{2}}A_{aI}T ^{I}_B{}^A\P ^B
\eea
\\ \\
Now, it is again straightforward to check that $\cg _I^{tot}$ and
$\tilde{\ch}_a^{tot}:=\ch _a^{tot} - A_a^I\cg _I^{tot}$ generate gauge
transformations and spatial diffeomorphisms, respectively. We also knows that
all constraints are diffeomorphism-covariant, and that $\cg _I^{tot}$ and
$\ch ^{tot}$ are gauge-covariant. This means again that the cruical calculation
is
$\{\ch ^{tot}[N],\ch ^{tot}[M]\}$. Here, however, we run into trouble; this
last
Poisson bracket fails to close the algebra. We get
\bea \{\ch ^{tot}[N],\ch ^{tot}[M]\}&=&\ch ^G_a[q^{ab}(N\partial _bM-M\partial
_bN) ]
+ \nn \\
&& \int _{\Sigma}d^3x\left ( \sqrt{2}(M\partial _aN - N\partial _aM)
E^{aI}E^{bJ}
\P ^A\cd _b\l _B  \times \right.
 \nn \\
&& \left.( \e _{IJK}({\scriptstyle E^{dL}})T ^K_A{}^B + \sqrt{2}T _{IA}{}^B
T_{JB}{}^D) \right ). \label{HH3} \eea
The reason why the constraint algebra closes when the Lie-algebra is $so(3)$
and the
$T _{IA}{}^B$'s are the Pauli-matrices, is that for that case $\e _{IJK}
({\scriptstyle E^{dL}})=f_{IJK}$ and we have the identity $T _{IA}{}^BT
_{JB}{}^C=-\frac{1}{2}\d _{IJ}\d _A^C - \frac{1}{\sqrt{2}}f_{IJ}{}^{K}T
_{KA}{}^C$,
meaning
that we get a cancelation of "bad" terms in (\ref{HH3}). Thus, although it is
possible
to find other Lie-algebras whose representations satisfy identities of the
above type,
it does not seem possible to cancel the term containing $\e _{IJK}
({\scriptstyle E^{dL}})$, for higher dimensional Lie-algebras. Therefore, it
seems
that the naive generalization of the conventional gravity-spinor coupling fails
to
give a closed constraint algebra for Lie-algebras of dimension $>3$.\\ \\
The solution to this problem might be that there exist another spinor coupling
that
gives a correct constraint algebra and reduces to the conventional one when the
Lie-algebra is chosen to be $so(3)$. We have, however, not managed to find such
a
coupling.

\section{Static, spherically symmetric solution to the $U(2)$
theory.}\label{solution}
In some cases, it is much more rewarding to study an explicit solution to a
theory than to study the full theory. It may, for instance, happen that some
unphysical features of the theory are completely obvious in the explicit
solution, while it is very hard to notice them in the full theory.

In this section, we will therefore derive the static and spherically
symmetric solution to the unified model, for gauge group $U(2)$, and compare
it to the normal Reissner-Nordstr\"{o}m solution (with a cosmological
constant). The Reissner-Nordstr\"{o}m solution is a static and spherically
symmetric solution to the conventional Einstein-Maxwell theory. In subsection
(\ref{ansatz}) we will give the ansatz for the static and spherically
symmetric fields, and also derive the quantities that are common for the
conventional and the unified theory. In subsections (\ref{conv}) and
(\ref{uni}) we derive the solutions for these two different theories.

\subsection{Static and spherically symmetric ansatz}\label{ansatz}
To find the spherically symmetric ansatz for the fields, we will use
 the results in \cite{another}. In this reference, one requires
that the Lie-derivative along the three rotational vector fields, when acting
on the fields, will correspond to a constant gauge transformation. That is
\be \pounds _{L^a}V=\{ V,\cg _i[\L ^i]\} \ee
where $L^a$ denote the three rotational vector fields, $V$ represent any
field in the theory, and $\L ^i$ is a constant gauge-parameter. By solving
these equations, an ansatz was found. We will use the same ansatz here, but
first we perform a gauge rotation to minimize the number of trigonometric
functions in the fields. By using the $SO(3)$ group element
\be G=\left (\begin{array}{ccc}\sin \q \cos \f & \sin \q \sin \f & \cos \q \\
\cos \q \cos \f & \cos \q \sin \f & -\sin \q \\ -\sin \f & \cos \f & 0
\end{array}\right ) \ee
we transform the ansatz in \cite{another}, and find the following expressions
\bea E^{r}_i&=&E_1 \sin \q V_i \\
E^\q _i&=&E_2\sin \q Z_i + E_3\sin \q W_i \\
E^\f _i&=&-E_2W_i + E_3Z_i \\
A_r^i&=&A_1V^i\\
A_\q ^i&=&\frac{1}{2}A_3W^i+\frac{1}{2}A_2Z^i\\
A_\f ^i&=&\cos \q V^i -\frac{1}{2}A_2\sin \q W^i + \frac{1}{2}A_3\sin \q
Z^i\eea
for the $so(3,C)$ valued fields $E^{ai}$ and $A_{ai}$. The index $r, \q $ and
$\f $ denote the normal spherical coordinates. The internal vectors $V^i,
W^i$ and $Z^i$ span a constant right-handed orthonormal basis for the internal
 space. For instance: $V^i=(1,0,0)$, $W^i=(0,1,0)$ and $Z^i=(0,0,1)$ is a
good choice. $E_1, E_2, E_3, A_1, A_2$ and $A_3$ are all functions of $r$.
When performing the gauge transformation on the connection
$A_{ai}$, one must first choose a three dimensional representation for the
Lie-algebra, use the transformation rule:
\be A_a^G=GA_aG^T-(\partial _aG)G^T \ee
and then go back to the vector representation.

Now, since the magnetic field has the same properties as the electric field
under Lie-derivative and gauge-transformations, it follows that $B^{ai}$ gets
the same ansatz as $E^{ai}$ above. By using the ansatz for $A_{ai}$,
the following relations can be derived:
\bea B_1&=&\frac{1}{2}(A_2^2+A_3^2-4) \\
B_2&=&A_1A_2-A_3'\\
B_3&=&A_2^\prime  +A_1A_3\eea
The prime denotes derivative w.r.t the $r$-coordinate. The other fields in
the theory are restricted by the spherical symmetry to have
 the following form:
\bea N&=&\frac{\tilde{N}(r)}{\sin \q} \\
N^a&=&(N^r(r),0,0)\\
A_{0i}&=&A(r)V_i\\
A_0&=&a(r)\\
E^a&=&\sin \q (E^r(r),0,0)\\
A_a&=&(A_r(r),0,0)\Rightarrow B^a=(0,0,0) \eea \\ \\
The two theories that will be compared here, are the conventional minimally
coupled Einstein-Maxwell theory and the unified theory for gauge group
$U(2)\cong SO(3)\times U(1)$. For the conventional theory, the gravity
Hamiltonian is given by (\ref{1}) and the Maxwell part by (\ref{YMHam}) with
gauge group $U(1)$. The Hamiltonian for the unified theory is given by
(\ref{uniH}). We notice that the only difference lies in the
Hamiltonian constraint, meaning that we can treat the other constraints
without specifying to which theory they belong. We list the result for the
$SO(3)$ and the $U(1)$ Gauss law, as well as the vector constraint:
\bea \cg_i&=&\partial _aE^a_i + f_{ijk}A_a^jE^{ak}=\sin \q V_i(E_1' + E_2 A_3
- A_2 E_3) \label{s1} \\
\cg&=&\partial _aE^a=E^{r\prime}  \label{s2} \\
\ch _a&=&\frac{1}{2}\e _{abc}(E^{bi}B^c_i+E^bB^c)=\d ^r_a\sin \q (E_2 B_3 -
E_3 B_2) \label{s3} \eea

\subsection{Conventional theory}\label{conv}
In the conventional theory, the Hamiltonian constraint is
\be \ch = -\frac{1}{4}\e _{abc}\e
_{ijk}E^{ai}E^{bj}(B^{ck}+\frac{2i\l}{3}E^{ck})+\frac{1}{2}\sqrt{\mid
q^{ab}\mid }q_{ab}(E^aE^b+\frac{1}{4}B^aB^b) \label{sH} \ee
where $q^{ab}:=-E^{ai}E^b_i$ and $q_{ab}$ is its inverse. We also know that
the densitized spacetime metric is given by
\be \tilde{g}^{\a \b}=\sqrt{-g}g^{\a \b}=\left (\begin{array}{cc}-\frac{1}{N}
& \frac{N^a}{N} \\ \frac{N^a}{N} & -N E^{ai}E^b_i -\frac{N^aN^b}{N}
\end{array} \right ) \label{sMetric}\ee
and its determinant is
\be \sqrt{-g}=N\sqrt{-\mid E^{ai}E^b_i\mid} \label{det} \ee
Note that we will assume the spatial metric $q^{ab}$ to be positive definite
and real. Using the ansatz given in the previous subsection, we get
\be \ch = \sin ^2\q \left ( \frac{i}{2}\frac{E_2^2+E_3^2}{E_1}(E^r)^2 -
\frac{1}{2}B_1(E_2^2+E_3^2)-E_1(E_3B_3+E_2B_2)-i\l E_1(E_2^2+E_3^2)\right )
\label{s8} \ee
Now it is straightforward to derive all the equations of motion, but before
we write out the result, we will fix the remaining gauges. It is four gauges
that need to be fixed: one component of the $SO(3)$ rotations, the radial
part of the spatial diffeomorphisms, the $U(1)$ symmetry, and the time-like
diffeomorphisms. The gauges we choose are
\bea A_1&=&0 \\
A_3&=&0 \\
A_r&=&0 \\
E_1&=&-ir^2 \eea
We will not go into details about these choices. It is, however,
straighforward to check that $A_1=0$ can be reached by an $SO(3)$
transformation from a generic field configuration. Similarly, $A_3=0$ and
$A_r=0$ can always be reached by transformations generated by $\ch $
and $\cg $, respectively. The reason for the choice $E_1=-i r^2$ is
that we want to recover the standard Reissner-Nordstr\"{o}m form of
the solution (in Schwarzschild coordinates). This choice is, however,
not completely generic. There is another gauge choice that gives a
physically inequivalent solution: $E_1=constant$. But since we are
only interested in comparing the standard Reissner-Nordstr\"{o}m type
solution, here, we will only study the former choice.

With these choices, the remaining equations are
\bea E^r&=&q=constant \\
E_2&=&0\\
N^r&=&0\\
A_2&=&-\frac{2ir}{E_3} \label{s35}\\
A_2^\prime  &=&-\frac{i}{2}(\frac{1}{2}A_2^2-2)\frac{E_3}{r^2}-i\l E_3
-\frac{i}{2}\frac{E_3q^2}{r^4}\label{s36}\\
A^\prime &=&\tilde{N}(E_3A_2^\prime   + i \l E_3^2 -\frac{i}{2}\frac{E_3^2q^2}
{r^4})
\label{s37} \\
A_2A&=&\tilde{N}(E_3(\frac{1}{2}A_2^2-2)-ir^2A_2^\prime +2\l r^2 E_3
+\frac{E_3q^2}{r^2})\label{s38} \\
A&=&\frac{\tilde{N}}{2}A_2E_3+\frac{i}{E_3}\partial _r(\tilde{N}r^2E_3)
\label{s38b}\\
a^\prime  &=&\frac{\tilde{N}E_3^2}{r^2}q\label{s34} \eea
Using (\ref{s35}) and (\ref{s36}) it is possible to solve for $E_3$ and $A_2$:
\bea E_3&=&-\frac{ir}{\xi (r)} \label{s39} \\
A_2&=&2\xi (r) \label{s40} \\
\xi (r)&:=&\sqrt{1-\frac{C}{r} - \frac{\l r^2}{3} + \frac{q^2}{2r^2}} \eea
where $C$ is a constant of integration.
With this solution at hand, (\ref{s37}) and (\ref{s38}) give $\tilde{N}$ and
$A$:
\bea \tilde{N}&=&D\frac{\xi ^2(r)}{r^2} \label{sN} \\
A&=&-\frac{D}{2}\partial _r\xi ^2(r) \label{s44} \eea
where $D$ is another constant of integration.
Now, using the expression for the metric (\ref{sMetric}), the line-element
becomes:
\be ds^2=-D\xi ^2(r)dt^2+\xi ^{-2}dr^2 + r^2d\q ^2 + r^2\sin ^2\q d\f ^2
\label{RN} \ee
which is the standard Reissner-Nordstr\"{o}m solution in Schwarzschild
coordinates. The conventional choices for $D$ and $C$ are: $D=1$ and $C=2M$.
The first choice just corresponds to a normalization of the time-coordinate,
and the second one is related to the fact that $M$ is the ADM-mass if $\l =0$.

\subsection{The unified $U(2)$ theory}\label{uni}
Since the gauge group $U(2)$ locally is isomorphic to $SO(3)\times U(1)$, we
choose to split the theory into these parts directly from the beginning. Here,
we have the problem of using a non
semi-simple group, meaning that the Cartan-Killing form is degenerate.
Therefore, we
will instead choose $g_{IJ}=\d _{IJ}$ as our bilinear form. Note, however,
that there is nothing that stops us from changing sign of the $U(1)$
component of this "group-metric". In the end, these choices are simply
related by an imaginary rescaling of the $U(1)$ fields.

The Hamiltonian for the unified theory is
\bea \ch &=&-\frac{1}{4}\e _{abc}\e_{IJK}({\scriptstyle E^{dL}})E^{aI}E^{bJ}
(B^{cK}+\frac{2i\l}{3} E^{cK})=\frac{i}{2}\sqrt{\mid q^{ab}\mid
}(q_{ab}E^{aI}B^b_I -2i\l )\nn \\ &=&\frac{i}{2}\sqrt{\mid q^{ab}\mid
}(q_{ab}E^{ai}B^b_i + q_{ab}E^aB^b - 2i\l ) \label{uH} \eea
where $q^{ab}=-E^{aI}E^b_I=-E^{ai}E^b_i-E^aE^b$ and $q_{ab}q^{bc}=\d _a^c$.
The $U(2)$ indices are denoted $I, J, K, ...$, and for $SO(3)$ we use $i, j,
k ...$.

The densitized spacetime metric is now given by
\be \tilde{g}^{\a \b}=\sqrt{-g}g^{\a \b}=\left (\begin{array}{cc}-\frac{1}
{N}
& \frac{N^a}{N} \\ \frac{N^a}{N} & -N E^{aI}E^b_I -\frac{N^aN^b}{N}
\end{array} \right ) \label{uMetric}\ee
and its determinant is
\be \sqrt{-g}=N\sqrt{-\mid E^{aI}E^b_I\mid}  \ee
To not get into trouble about the double-valueness of the square root for
complex fields, we will always arrange the argument of the square root to be
real and positive definite. Then, we choose the positive branch. The other
branch just corresponds to $N \rightarrow -N$.

With the static, spherically symmetric ansatz given in section \ref{ansatz},
the
Hamiltonian constraint becomes:
\be \ch = \frac{i}{2}\sqrt{-(E_1^2+(E^r)^2)}(E_2^2+E_3^2)\sin ^2\q\left (
\frac{E_1B_1}{E_1^2+(E^r)^2} + 2\frac{E_2B_2+E_3B_3}{E_2^2+E_3^2} + 2i\l
\right ) \label{s49} \ee
Now, it is again straightforward to derive the equations of motion from
(\ref{s1})-(\ref{s3}), (\ref{s49}),
 but before writing them out, we fix the remaining
 gauges:
\be A_1=0\hspace{20mm} A_3=0 \hspace{20mm} E_1=-i\sqrt{r^4 +
(E^r)^2}\hspace{20mm} A_r=0 \ee
The reason for these choices are the same as for the conventional theory: the
two first can always be reached from a generic field-configuration with
transformations generated by the Gauss law and the Hamiltonian constraint,
respectively. The third choice is due to the wish to recover the solution in
Schwarzschild-like coordinates. Note, however, that this choice is not
completely generic. There exist another choice $E_1=const$ that gives a
non-equivalent solution.

With these gauge choices, the constraints and the equations of motion become:
\bea E_2&=&0 \\
N^r&=&0 \\
E^r&=&\tilde{q}=constant \\
A_2&=&-\frac{2ir^3}{E_3\sqrt{r^4+\tilde{q}^2}} \label{s76} \\
A_2^\prime &=&
\frac{iE_3}{2}\frac{\sqrt{r^4+\tilde{q}^2}}{r^4}(\frac{1}{2}A_2^2-2)-iE_3\l
\label{s77} \\
A^\prime &=& -\frac{i\tilde{N}}{2}\frac{2\tilde{q}^2+r^4}{r^6}E_3^2
(\frac{1}{2}A_2^2-2)\label{s78} \\
A_2A&=& i\tilde{N}r^2A_2^\prime \label{s79} \\
A&=& \frac{\tilde{N}}{2}\frac{\sqrt{r^4+\tilde{q}^2}}{r^2}E_3A_2
+\frac{i}{E_3}\partial _r(\tilde{N}r^2E_3) \label{s80} \eea
Using (\ref{s76}) and (\ref{s77}), $E_3$ and $A_2$ are easily determined:
\bea E_3&=&-\frac{ir^3}{\sqrt{r^4+\tilde{q}^2}}\frac{1}{\tilde{\xi}(r)}
 \label{s81} \\
A_2&=& 2 \tilde{\xi}(r) \label{s82} \eea
where we have defined
\bea \tilde{\xi}(r)&:=&\sqrt{1-\frac{\tilde{C}}{r}-\frac{1}{r}I(r)} \\
I(r)&:=&\int ^rds\frac{\l s^4}{\sqrt{s^4+\tilde{q}^2}} \eea
and $\tilde{C}$ is a constant of integration.
Then, (\ref{s79}) and (\ref{s80}) give us $\tilde{N}$ and $A$:
\bea \tilde{N}&=&\frac{
\tilde{D}\sqrt{r^4+\tilde{q}^2}\tilde{\xi}^2(r)}{r^4} \\
A&=&\frac{i\tilde{D}}{2}\frac{\sqrt{r^4+\tilde{q}^2}}{r^2}\partial _r
\tilde{\xi}^2(r) \label{s86} \eea
where $\tilde{D}$ is another constant of integration.
Finally, one must check (\ref{s78}), which is satisfied with the solution
given above.

Altogether, we now get the line-element
\be
ds^2=-\tilde{D}\tilde{\xi}^2(r)dt^2+\frac{r^4}{r^4+\tilde{q}^2}\tilde{\xi}
^{-2}(r)dr^2+r^2d\q ^2 + r^2\sin ^2\q d\f ^2 \label{ds2} \ee
Far away from the charge, where $r^2\gg \tilde{q}$, we can expand $g_{tt}$
and $g_{rr}$:
\bea g_{tt}&=&-\tilde{D}\left (1-\frac{\tilde{C}}{r}-\frac{\l r^2}{3} -
\frac{\l
\tilde{q}^2}{2r^2} + \co (\frac{\tilde{q}^4}{r^8})\right ) \\
g_{rr}&=&\left (1-\frac{\tilde{C}}{r}-\frac{\l r^2}{3} - \frac{5\l
\tilde{q}^2}{6r^2} + \co (\frac{\tilde{q}^4}{r^8})\right )^{-1} \eea
Similarly, we may expand the metric close to the charge $r^2\ll \tilde{q}$
\bea  g_{tt}=-\tilde{D}\left ( 1-\frac{\tilde{C}}{r} -\frac{\l
r^4}{5\tilde{q}} + \co (\frac{r^6}{\tilde{q}^3})\right ) \\
g_{rr}=\frac{r^4}{\tilde{q}^2}(1-\frac{r^4}{\tilde{q}^2})\left (
1-\frac{\tilde{C}}{r} -\frac{\l
r^4}{5\tilde{q}} + \co (\frac{r^6}{\tilde{q}^3})\right )^{-1} \eea
Thus, we see that the solution coincides with the conventional
Schwarzschild-de Sitter solution when $\tilde{q}=0$, and it resembles the
Reissner-Nordstr\"{o}m solution for $r^2\gg \tilde{q}$. Note, however, that
the numerical factors in the electric charge-terms do not agree. Remember
also that the physical charge is $q_{phys}\sim \sqrt{\l}\tilde{q}$. We also
see
 that the wormhole like feature of the solution that was found in the
(2+1)-dimensional
case \cite{2+1} does not appear here. That is because the sign inside the
square root $\sqrt{r^4+\tilde{q}^2}$ is a plus-sign, whereas the
corresponding square root in (2+1)-dimensions had a minus-sign. This sign is,
however, closely related to the choice of bilinear invariant form of the
Lie-algebra, and by choosing this bilinear form to be $g_{IJ}=diag(1,1,1,-1)$
we would have recovered the same solution with $\tilde{q}^2\rightarrow
-\tilde{q}^2$. In the true theory, this sign will depend on how the $U(1)$
and $SO(3)$ algebras are embedded in the full algebra.
We have not tried to extend this solution beyond the horizons.

\section{Conclusions}
We conclude this paper by noting that the unified model presented here is
interesting mainly because it gives the conventional Yang-Mills theory in the
weak field limit. However, this theory still has a few unwanted features, as
{\it e.g} the need for complex fields for Lorentzian spacetime, and
the incapability
of selecting a unique gauge group. Therefore, we do not believe that this
model is to be taken serious as a candidate theory for the unified
description of gravity and Yang-Mills theory. Instead, we hope for a
modification of this theory that will solve the above mentioned problems.\\ \\

{\bf Acknowledgements}\\ \\
We thank Ingemar Bengtsson and Fernando Barbero for discussions and
suggestions.
 S.C is grateful to U.G.C (India) and Council
for International Exchange of
Scholars (CIES), for the opportunity to visit the Center for
Gravitational Physics and Geometry, Penn State University. S.C's work was
supported in part by the grant from CIES, grant no. 17263. P.P's work was
supported by the NFR (Sweden) contract no. F-PD 10070-300, and {\it Per Erik
Lindahls
Fond, Kungliga Vetenskapsakademien}.

\newpage
\appendix
\section{Legendre transforms}\label{Legendre}
Here, we will derive the Hamiltonian formulation of the Yang-Mills Lagrangian.

The Lagrangian density for the Yang-Mills theory is
\be \cl ^{YM}=\frac{k}{4}\sqrt{-g}g^{\a \b}g^{\d \e}F_{\a \d}^AF_{\b \e A} \ee
where $g^{\a \b}$ is the inverse metric, $g$ the determinant of the metric,
and $F_{\a \b}^A$ is the YM field strength in the vector representation. We
also have an arbitrary multiplicative constant $k$ included.
First, we do a (3+1)-decomposition:
\be \cl ^{YM}= \frac{k\sqrt{-g}}{2}\left ( g^{00}g^{ab}F_{0a}^AF_{0bA} -
g^{0a}g^{0b}F_{0a}^AF_{0bA} + 2 g^{0a}g^{bc}F_{0b}^AF_{acA} +
\frac{1}{2}g^{ab}g^{cd}F_{ac}^AF_{bdA}\right ) \ee
and then define the momenta canonically conjugate to the YM connection
$A_{aA}$:
\be E^a_A:=\frac{\d L^{YM}}{\d \dot{A}_a^A}=k\sqrt{-g}\left (
g^{00}g^{ab}F_{0bA}-g^{0b}g^{0a}F_{0bA} + g^{0b}g^{ac}F_{bcA}\right )
\label{YME} \ee
Now we introduce the ADM-decomposition of the metric:
\bea g^{00}&=&-\frac{1}{M^2}\hspace{25mm} g^{0a}=\frac{N^a}{M^2} \nn \\
g^{ab}&=&h^{ab}-\frac{N^aN^b}{M^2} \Rightarrow \sqrt{-g}=\frac{M}{\sqrt{
\mid h^{ab}\mid }} \eea
Then we invert the relation (\ref{YME}):
\be F_{ob}^A=-\frac{M\sqrt{\mid h^{ab}\mid}}{k}h_{bc}E^{cA} + \frac{1}{2} N^e
\e _{ebd}B^{dA} \label{F0} \ee
where $h_{ab}$ is the inverse to $h^{ab}$, and we have defined the YM
magnetic field: $B^{aA}:=\e ^{abc}F_{bc}^A$. Thus, the total YM Hamiltonian
density becomes
\bea \ch ^{tot}&=&E^a_A\dot{A}_a^A-\cl ^{YM}=E^a_A\left (F_{0a}^A+\cd
_aA_0^A\right ) -\cl ^{YM}
=-\frac{M\sqrt{\mid h^{ab}\mid}}{2k}h_{ab}E^{aA}E^b_A \nn
\\ &&-\frac{kM\sqrt{\mid
h^{ab}\mid}}{8}h_{ab}B^{aA}B^b_A + \frac{1}{2}N^a\e _{abc}E^{bA}B^c_A -
A_{0A}\cd _aE^{aA} \label{YM1} \eea
Finally, to make contact with the Ashtekar variables, we define
\be q^{ab}:=-E^{ai}E^b_i:=\frac{h^{ab}}{\mid h^{cd}\mid}\hspace{25mm}
N:=M\sqrt{\mid h^{ab}\mid} \ee
which gives the final Hamiltonian
\bea H^{YM}&=&\int _{\Sigma}d^3x \left (N\ch ^{YM} + N^a\ch _a^{YM} -A_{0A}
\cg ^A_{YM}\right ) \label{YMHam} \\
\ch ^{YM}&=& -\frac{1}{2}\sqrt{\mid q^{ab}\mid}q_{ab}\left (
\frac{1}{k}E^{aA}E^b_A +
\frac{k}{4}B^{aA}B^b_A\right ) \\
\ch _a^{YM}&=&\frac{1}{2}\e _{abc}E^{bA}B^{c}_A \\
\cg ^{YM}_A&=& \cd _aE^a_A=\partial _aE^a_A + f_{ABC}A_a^BE^{aC} \eea
If one picks an invariant bilinear form for the YM Lie-algebra that is
positive definite, one normally considers a negative $k$ in order to get a
positive definite Hamiltonian.

\section{Dimensions and units}\label{units}
Here, we want to introduce dimensions and a unit-system in the theory
discussed in this paper. The normal convention to put some or all of the
fundamental constants of nature equal to one, is very useful in doing
calculations. However, a soon as one wants to start comparing with reality,
it is often more convenient to express the results in a system of units that
one is used to. We have chosen to use the SI unit-system here. In this
system, the fundamental units for length, mass, time and charge are meter
(m), kilogram (kg), second (s) and Coulomb (C). (Charge is equal to current
times time, and often it is convenient to use Ampere (A) instead of Coulomb
as a fundamental unit. The following is true: 1 C=1 As). To introduce units
in the theory, we need a set of dimension-full constants: the gravitational
constant, $G=6.67\times 10^{-11}\frac{m^3}{s^2kg}$, the speed of light,
$c=3\times 10^{8}\frac{m}{s}$, the Planck constant, $\hbar=1.05\times
10^{-34}\frac{kg\:m^2}{s}$. When using the SI-units, one also needs the
dielectric constant for vacuum, $\e _0=8.85\times
10^{-12}\frac{A^2s^4}{kg\:m^3}$.\\ \\
Now, we know that a good action should have dimension {\rm energy$\times
$time}, which has units: $[S]=\frac{kg\:m^2}{s}$ Here, the square brackets
denote the units of the enclosed object. $S$ stands for the action. If we
then assume that all our spacetime coordinates have dimension length (we
choose a coordinate system where this is true), the metric becomes
dimension-less, and the Riemann tensor gets dimension $length^{-2}$.
Consequently, the Einstein-Hilbert action for gravity gets dimension
$length^2$:
\be \begin{array}{cccc}[g_{\a \b}]=1 & [R_{\a \b \d}{}^{\e}]=\frac{1}{m^2} &
[d^4x]=m^4\Rightarrow & [S_{EH}]= [d^4x][\sqrt{-g}][R]=m^2
\end{array}\ee
To get the correct dimension on the action, we need to multiply
it with a constant of dimension $\frac{mass}{time}$. The
combination $\frac{c^3}{G}$ will do the job, meaning that the
dimensionally correct action is (throughout this paper, we have neglected the
conventional factors of $\pi$)
\be S=\frac{c^3}{G}\int _{\cm}d^4x\sqrt{-g}R \ee
To find out the correct dimension for the Maxwell field, $A_{\a}$, we start
from the expression of the Lorentz-force on a charged particle in an
electromagnetic field:
\be \vec{F}=Q(\vec{E}+\vec{v}\times \vec{B}) \ee
Here, $Q$ is the charge, $\vec{F}$ the force, $\vec{E}$ the electric field,
$\vec{v}$ the
velocity, and $\vec{B}$ the magnetic field. Since the unit for force is
$1N=1\frac{kg\:m}{s^2}$, it follows that the units for the electromagnetic
fields are
$[\vec{B}]=\frac{kg}{A\:s^2}$ and $[\vec{E}]=\frac{kg\:m}{A\:s^3}$. With
$\vec{B}=\bigtriangledown \times \vec{A}$, the unit for the vector potential
becomes: $[A_\a ]=[\vec{A}]=\frac{kg\:m}{A\:s^2}$ (To make contact with the
conventional units for measuring electromagnetic fields, one can use the
relations: $1V=\frac{kg\:m^2}{A\:s^3}$ and $1T=\frac{kg}{A\:s^2}$.)
Now, it is an easy task to calculate the units of the Maxwell action:
\be [S_{Max}]=[d^4x][\sqrt{-g}][g^{\a \b}g^{\d \e}][F_{\a \d}F_{\b
\e}]=\frac{kg^2m^4}{A^2s^4} \ee
Thus, we need to multiply this action with a constant with units
$\frac{A^2s^3}{kg\:m^2}$, and the natural choice for this constant is $\e
_0c$. so, the total action for the coupled Einstein-Maxwell theory becomes
\be S^{tot}= \frac{c^3}{G}\int _{\cm}d^4x\sqrt{-g}(R-2\l)-\e _0c\int
_{\cm}d^4x\sqrt{-g}g^{\a \b}g^{\d \e}F_{\a \d}F_{\b \e} \ee
Our conventions for the gravitational fields are that we keep all metric
variables ($E^{ai}, N, N^a$) dimension-less, and let the connection have
dimension $length^{-1}$. By doing that, we do not need to introduce $G$ into
the pure gravity theory. Effectively this means that we have pulled out the
factor $\frac{G}{c^3}$ from the total action, and we instead get a factor
$\frac{\e _0G}{c^2}$ multiplying the Maxwell action. And, since the
Hamiltonian formulation of the Maxwell (Yang-Mills) action is found without
this factor, we must introduce this factor in the Hamiltonian constraint
according to (\ref{YMHam}):
\be \ch ^{Max}=\sqrt{\mid q^{ab}\mid}q_{ab}(\frac{c^2}{\e
_0G}E^aE^b +\frac{1}{4}\frac{\e _0G}{c^2}B^aB^b) \ee
(To further complicate things, note that the physical electric field really
is: $E^a=\frac{\e _0G}{c^3}E^a_{phys}$ due to the rescaling of the momenta in
the Legendre transform (\ref{YME})) Now, it is straightforward to compare this
Hamiltonian with the result of the weak-field expansion of the unified model
(\ref{40}). Thus, we see that
\bea \sqrt{\l}\tilde{e}^a &\sim &\sqrt{\frac{c^2}{\e _0G}}E^a=\sqrt{\frac{c^2}
{\e _0G}}\frac{\e _0G}{c^3}E^a_{phys} \\
\frac{1}{\sqrt{\l}} b^a &\sim &\sqrt{\frac{\e _0G}{c^2}}B^a_{phys} \eea
meaning that
\bea \d ^{ab} & \gg &\tilde{e}^a\tilde{e}^b\sim \frac{\e
_0G}{c^4\l}E^a_{phys}E^b_{phys} \\
\l ^2 \d ^{ab} & \gg & b^ab^b\sim \frac{\e _0G\l}{c^2}B^a_{phys}B^b_{phys} \eea
Numerically this means: $E^a_{phys}\ll 10^{-4}\frac{V}{m}$ and
$B^a_{phys}\ll 10^{-12}T$! Where we have used the value $10^{-62}\:m^{-2}$
\cite{300} for the cosmological constant. Note that the restriction on the
electromagnetic fields scales as $\sqrt{\l}$.\\ \\
Finally, we want to say a few word about the coupling to spinor fields.
Normally when one couples spinors to electromagnetism, one uses the minimal
coupling
\be \cd _\a\J=\partial _a\J + i\:e\:k\:A_{\a}\J \label{hbar} \ee
where $e$ is the charge of the electron an $k$ is some dimension-full constant
used to get consistent dimensions for both terms. Now, by using the units for
$A_{\a}$ given above $[A_{\a}]=\frac{kg\:m}{A\:s^2}$, we see that $k$ must
have units: $[k]=\frac{s}{kg\:m^2}$, the units of $action^{-1}$. The only
reasonable choice for such a constant is $\frac{1}{\hbar}$. At first sight it
seems strange that one has to introduce Planck's constant already in the
classical theory, however, one may also have the viewpoint that the concept
of spinors
naturally belongs to the quantum theory anyway. That is, it does not make
much sense to talk about a classical spinor. Furthermore, when one couples a
charged point-particle to electromagnetism, the term corresponding to
(\ref{hbar}) is really: $p_{\a}-e A_{\a}$, which in the quantization
introduces the $\hbar$ by the definition $\hat{p}_{\a}=i\:\hbar\partial
_{\a}$.

However, when we introduce the $U(1)$ covariant derivative
 in the unified model, we do not have any elementary charge or
Planck constant available. Instead, combinations of the other constants will
appear in the coupling. From (\ref{covder}) we see that
\be \cd _a\J=\partial _a\J +i\: A_a\J \ee
for the $U(1)$ connection in the unified model, and we also know the relation
between this connection and the physical Maxwell connection:
\be A_a \sim \sqrt{\frac{\l \e _0 G}{c^2}} A_a^{phys} \ee
Hence, the coupling constant $\frac{e}{\hbar}$ has been replaced by $
\sqrt{\frac{\l \e _0 G}{c^2}}$. These two constants do not numerically agree
(unless the cosmological constant is extremely large), but that is really
irrelevant since the value of this constant is renormalized in the quantum
theory. \\ \\

\end{document}